\documentclass[final]{eps}

\usepackage{graphicx}
\usepackage{times}

\volumenumber{63}
\publishyear{2011}
\frompage{\pageref{firstpage}}
\topage{\pageref{finalpage}}

\shorttitle{A. K. Inoue: The origin of dust in galaxies}

\title{The origin of dust in galaxies revisited: the mechanism
determining dust content}
\author{Akio K. Inoue$^1$}
\affiliation{$^1$College of General Education, Osaka Sangyo University, 
3-1-1 Nakagaito, Daito, Osaka 574-8530\\}
\received{xxxx xx, 2010}\revised{xxxx xx, 2010}\accepted{xxxx xx, 2010}
\abstract{The origin of cosmic dust is a fundamental issue in planetary
science. This paper revisits the origin of dust in galaxies, in
particular, in the Milky Way, by using a chemical evolution model of a
galaxy composed of stars, intestellar medium, metals (elements
heavier than helium), and dust. We start from a review of 
time-evolutionary equations of the four components, and then, we 
present simple recipes for the stellar remnant mass and yields of metal
and dust based on models of stellar nucleosynthesis and dust
formation. After calibrating some model parameters with the data from the
solar neighbourhood, we have confirmed a shortage of the stellar
dust production rate relative to the dust destruction rate by supernovae
if the destruction efficiency suggested by theoretical works is
correct. If the dust mass growth by material accretion in
molecular clouds is active, the observed dust amount in the solar
neighbourhood is reproduced. We present a clear analytic
explanation of the mechanism for determining dust content in galaxies
after the activation of accretion growth: a balance between accretion
growth and supernova destruction. Thus, the dust content is
independent of the uncertainty of the stellar dust yield after
the growth activation. The timing of the activation is determined by 
a critical metal mass fraction which depends on the growth and
destruction efficiencies. The solar system formation seems to have
occured well after the activation and plenty of dust would have existed 
in the proto-solar nebula.}

\begin{document}
\label{firstpage}
\maketitle
\copyrighttext{}

\section{Introduction}

Cosmic dust grains are negligible in mass in the Universe. 
Nevertheless, they play significant roles on a lot of astronomical,
astrophysical, and astrochemical aspects: extinction (absorption
and scattering) matter of radiation, an emission source in
infrared wavelengths, a coolant and a heat source in the
interstellar medium (ISM) and intergalactic medium (IGM), and 
a site of formation of molecules.
Therefore, dust is one of the most important ingredients in the
Universe. Dust is also important for planetary science
because grains are material for planets.

Dust grains are formed in rapidly cooling gas of stellar outflows
(Yamamoto \& Hasegawa 1977; Draine \& Salpeter 1977). We call such
grains `stardust'. Sources of the stardust are asymptotic giant
branch (AGB) stars, supernovae (SNe), red supergiants, novae, Wolf-Rayet
stars, and so on (e.g., Gehrz 1989). The main source of the stardust in
the present Milky Way and the Magellanic Clouds is thought to be AGB 
stars (Gehrz 1989; Draine 2009; Matsuura et al.~2009).

SNe may also produce a significant amount of stardust (Kozasa \&
Hasegawa 1987; Todini \& Ferrara 2001; Nozawa et al.~2003, 2007;
Schneider et al.~2004; see also Kozasa et al.~2009). Stardust from
SNe is particularly important in the early Universe because the time for
stars to evolve to the AGB phase is typically about 1 Gyr, but the
cosmic time in the early Universe is shorter than it (Morgan \& Edmunds
2003; Maiolino et al.~2004; Dwek et al.~2007; but see also Valiante et
al.~2009). The `first' stardust may also play an important role to
change the mode of star formation from massive star dominated to
present-day Sun-like star dominated (Schneider et al.~2003, 2006).

However, dust formation by SNe remains in controversy
observationally. First detections of a few $M_\odot$ dust freshly
formed, which is much larger than expected, in Cassiopeia A (Cas A) and
Kepler SN remnants (SNRs) by submilimeter observations with {\it SCUBA}
(Dunne et al.~2003; Morgan et al.~2003) were almost contaminated by
foreground dust in the ISM on the sight-lines (Krouze et al.~2004; Gomez
et al.~2009). Recent infrared observations with {\it Spiter Space
Telescope} and {\it AKARI} and submilimeter observations with {\it
Herschel} and {\it BLAST} of Cas A and other SNRs are agreeing
with theoretical expectations of 0.01--0.1 $M_\odot$ per one SN (Rho et
al.~2008; Sakon et al.~2009; Nozawa et al.~2010; Barlow et al.~2010;
Sibthorpe et al.~2010).

Once stardust grains are injected into the ISM, they are processed
there. The grains in hot gas are bombarded by thermally moving protons
and sputtered (Onaka \& Kamijo 1978; Draine \& Salpeter 1979). SN
shock waves probably destroy dust grains by grain--grain collisional
shattering as well as sputtering (e.g., Dwek \& Arendt~1992; Jones
et al.~1994, 1996; Nozawa et al.~2006; Silvia et al.~2010). This
destruction process is widely accepted and observational evidences
of the destruction have been found in several SNRs, especially with {\it
Spitzer Space Telescope} (Arendt et al.~1991, 2010; Borkowski et al.~2006;
Williams et al.~2006; Dwek et al.~2008; Sankrit et al.~2010; but see
Mouri \& Taniguchi 2000).

Assuming the destruction efficiency expected by theoretical works,  
we obtain the life-time of dust grains of the order of 100 Myr 
(McKee 1989; Draine 1990; Jones et al.~1994, 1996). On the other hand,
the injection time of stardust is of the order of 1 Gyr (e.g.,
Gehrz 1989). Thus, another efficient channel of dust formation is
required to keep dust content in galaxies. The most plausible
mechanism is the accretion growth in the ISM (Draine 1990, 2009);
in dense molecular clouds, atoms and molecules of some refractory
elements and compounds accrete onto pre-existent grains and may change
from the gas phase to the solid phase. Note that unlike the
sticking growth of grains well studied in protoplanetary disks, this
accretion growth causes an increase in dust mass. This type of growth 
is favored to explain the observed depletions of some elements in the 
gas phase of the ISM relative to the solar abundance. The
correlation between the depletion degree and the density in the ISM is
particularly suggestive for this process (e.g., Savage \& Sembach 1996;
Jenkins 2009). It is also suggested that an efficient growth is
required to explain massive dust mass observed in the early Universe  
(Micha\l owski et al.~2010).

Since the pioneering work by Dwek \& Scalo (1980), many theoretical
works on dust content evolution in galaxies have been made so far
(Dwek 1998; Lisenfeld \& Ferrara 1998; Edmunds \& Eales 1998; Hirashita
1999a,b,c; Edmunds 2001; Hirashita et al.~2002; Inoue 2003; Morgan
\& Edmunds 2003; Dwek et al.~2007; Zhukovska et al.~2008; Calura et
al.~2008; Valiante et al.~2009; Pipino et al.~2011; Dwek \&
Cherchneff 2011; Gall et al.~2011a,b; Mattsson~2011; Asano et
al.~2011). These works are based on the evolutionary model of elemental
abundance in galaxies called chemical evolution model (Tinsley 1980 for
a review) and add some (or all) of the three processes of formation,
destruction, and growth of dust to it. One of the main results from
the recent works is the importance of the accretion growth.

This paper presents a new interpretation of the mechanism for
determining dust content in galaxies. Previous works imply that the
mechanism is a balance between dust destruction by SNe and accretion
growth in the ISM. However, this point has not been discussed clearly,
in contrast, this paper analytically shows that it is. For this aim, 
a simple one-zone model is sufficient. In addition, we present new
simple recipes describing stellar remnant mass and yields of elements
and dust from state-of-the-art models of stellar nucleosynthesis and
formation of stardust.

We will start from a review of basic equations presented in
\S2. In \S3, we present new simple recipes of stellar remnant mass and
yields. In \S4, we calibrate some model parameters to reproduce the
observed properties of the solar neighborhood. In \S5, we present our
analytical interpretation of the mechanism for determining dust
content in galaxies. We will present some further discussions in \S6. 
Experts of this field may go straight to \S5 which is the new
result of this paper.

Throughout this paper, we call elements heavier than helium `metal'
according to the custom of astronomy. We adopt the metal mass fraction
(so-called metallicity) in the Sun of $Z_\odot=0.02$ (Anders \& Grevesse
1989) conventionally, although the recent measurements suggest a smaller
value of 0.0134 (Asplund et al.~2009).

\section{Chemical and dust evolution model of galaxies}

\subsection{Equations of chemical and dust amount evolution}

We deal with a galaxy composed of stars (including their remnants;
i.e. white dwarfs, neutron stars, and black-holes) and the ISM. For
simplicity, we assume the ISM to be one-zone. The ISM contains metal and
dust as internal components. If we denote masses of these components as
$M_*$ (stars [and remnants]), $M_{\rm ISM}$ (ISM), $M_Z$ (metal), and
$M_{\rm d}$ (dust), the equations describing their time evolutions are  
(e.g., Dwek 1998)
\begin{equation}
 \frac{dM_*}{dt} = S(t) - R(t)\,,
\end{equation}
\begin{equation}
 \frac{dM_{\rm ISM}}{dt} = -S(t) + R(t) + I(t) - O(t)\,,
\end{equation}
\begin{equation}
 \frac{dM_Z}{dt} = -Z(t)S(t) + Y_Z(t) + I_Z(t) - O_Z(t)\,,
\end{equation}
\begin{equation}
 \frac{dM_{\rm d}}{dt} = -Z_{\rm d}(t)S(t) + Y_{\rm d}(t)
  - D_{\rm SN}(t) + G_{\rm ac}(t) + I_{\rm d}(t) - O_{\rm d}(t)\,,
\end{equation}
where $S$ is the star formation rate, $R$ is the mass return rate
from dying stars, $Y_Z$ and $Y_{\rm d}$ are the metal and dust 
supplying rates `yields' by dying stars, respectively. 
$Z\equiv M_Z/M_{\rm ISM}$ is the metal mass fraction in the ISM called 
`metallicity', and $Z_{\rm d}\equiv M_{\rm d}/M_{\rm ISM}$ is the dust
mass fraction in the ISM which we call the dust-to-gas mass ratio. Note
that $M_{\rm ISM}>M_Z \geq M_{\rm d}$. 

$I$, $I_Z$, and $I_{\rm d}$ are the ISM, metal, and dust infall rates
from the IGM, respectively. $O$, $O_Z$, and $O_{\rm d}$ are the ISM,
metal, and dust outflow rates to the IGM, respectively. In this paper,
we do not consider any outflows ($O=O_Z=O_{\rm d}=0$), but consider only
an ISM infall $I$ (no metal and dust in infalling gas: 
$I_Z=I_{\rm d}=0$), which is required to reproduce the metallicity
distribution of stars nearby the Sun.\footnote{$^1$Without gas infall
from the intergalactic space, we expect a much larger number of
low-metallicity stars at the solar neighborhood than the observed. This
is called the `G-dwarf problem' (e.g., Pagel 1989).} The reason why
we omit any outflows is that we do not know the transport mechanism of
metal and dust from galaxies to the IGM (e.g., Bianchi \& Ferrara
2005). However, this omission may be inconsistent with detections of
metal and dust in the IGM (e.g., Songaila \& Cowie 1996; M{\'e}nard et
al.~2010).\footnote{$^2$The origin of intergalactic metals and dust
is galactic outflows and the amount ejected from galaxies is the same
order of that remained in galaxies (e.g., M{\'e}nard et al.~2010 for
dust; see also Inoue \& Kamaya 2003, 2004, and 2010). Dust grains may be
ejected from galaxies more efficiently than metals because the grains
receive momentum through radiation pressure (Bianchi \& Ferrara 2005). 
Even in this case, our discussion about the dust-to-metal ratio in \S5
would not be affected essentially by omission of this selective removal
of dust, although the set of model parameters which can reproduce the
observations would change. In any case, this point would be an
interesting future work.}

In the dust mass equation (eq.~[4]), there are two additional
terms; $D_{\rm SN}$ is the dust destruction rate by SNe and $G_{\rm ac}$
is the dust growth rate in the ISM by metal accretion. These two
terms are discussed in \S2.5 and \S2.6 in detail.

\subsection{Star formation and infall rates}

We adopt a simple recipe for star formation introduced by Schmidt
(1959): $S\propto {M_{\rm ISM}}^p$ (Schmidt law). The index $p$ called
Schmidt index is observationally indicated to be $p=1$--2 (e.g., 
Kennicutt 1998; Elmegreen 2011) and some theoretical 
interpretations for the value are presented (e.g., Dopita \& Ryder
1994). However, the value and its origin of the index is still an
open problem (Elmegreen 2011 and references therein). Fortunately, the
choice of the index is not important in fact because in \S4 we calibrate
other model parameters so as to reproduce the observed star formation
history $S(t)$ at the solar neighborhood which is essential. We here
assume $p=1$ in order to solve the equations analytically in \S5.
In this case, we need a time-scale to give the star formation rate: star
formation time-scale, $\tau_{\rm SF}$ (see Table 1 in \S4 for the
values). Thus, the star formation rate is given by 
\begin{equation}
 S(t)=\frac{M_{\rm ISM}(t)}{\tau_{\rm SF}}\,.
\end{equation}

The infall from the IGM mimics the structure formation in the Universe
based on the hierarchical scenario with cold dark matter (e.g., Peacock
1999); small galaxies are first formed at density peaks of the dark
matter distribution in the Universe and they grow up larger and larger
as they merge each other and also obtain mass by accretion process.
Here we simply assume a smooth exponential infall rate although the mass
assembly of a galaxy is intrinsically episodic due to the merging
process. This simplification is a kind of ensemble average of many
galaxies and appropriate to examine a mean property of the galaxies.
The infall rate which we adopt is 
\begin{equation}
 I(t)=\frac{M_{\rm total}}{\tau_{\rm in}}\exp(-t/\tau_{\rm in})\,,
\end{equation}
where $\tau_{\rm in}$ is the infall time-scale and $M_{\rm total}$ is
the total mass which a galaxy obtains within the infinite time (see
Table 1 in \S4 for the values). Note that $M_{\rm total}$ just gives
the normalization of mass of a galaxy.

\subsection{Stellar mass spectrum and returned mass rate}

Salpeter (1955) first investigated the mass
spectrum of stars in the solar neighborhood, corrected it for the
modulation by stellar evolution and death, and obtained the mass
spectrum of stars when they are born, called initial mass function
(IMF) of stars. The Salpeter's IMF is a power-law: 
$dN/dm=\phi(m)\propto m^{-q}$ with $q=2.35$. A lot of following
researches confirmed that the slope was quite universal, especially for
massive stars, although there was a cut-off mass for low mass stars
(e.g., Kroupa 2002, Chabrier 2003 for reviews). We adopt here a simple
functional from proposed by Larson (1998) which is essentially
equivalent to the IMFs by Kroupa (2002) and Chabrier (2003) as 
\begin{equation}
 \phi(m)\propto m^{-q}\exp(-m_{\rm c}/m)\,,
\end{equation}
with a cut-off mass $m_{\rm c}$ and the range from $m_{\rm low}$ to 
$m_{\rm up}$. As a standard case, we adopt $p=2.35$, 
$m_{\rm c}=0.2$ $M_\odot$, $m_{\rm low}=0.1$ $M_\odot$, and 
$m_{\rm up}=100$ $M_\odot$. The cut-off mass well matches with the
observed data compiled by Kroupa (2002). We normalize the IMF as
$\int_{m_{\rm low}}^{m_{\rm up}}m\phi(m)dm=1$.

The mass returned rate from dying stars, $R$, is given by 
\begin{equation}
 R(t) = \int_{m_{\rm lf}(t)}^{m_{\rm up}} 
  \{m-w(m,Z[t'])\}\phi(m) S(t') dm\,,
\end{equation}
where 
\begin{equation}
 t'=t-\tau_{\rm lf}(m)
\end{equation}
is the time at which stars with mass $m$ dying at time $t$ are born, 
$\tau_{\rm lf}(m)$ is the stellar life-time, $w(m,Z)$ is the remnant
mass of stars with mass $m$ and metallicity $Z$, and $m_{\rm lf}(t)$ is
the minimum mass of stars dying at time $t$. This is the inverse
function of $t=\tau_{\rm lf}(m)$. If time $t$ is less than the life-time
of the star with $m_{\rm up}$, the returned rate $R=0$. We have assumed
that the metallicity of a star is the same as the ISM metallicity at the
time when the star is born.

The stellar life-time $\tau_{\rm lf}(m)$ is calculated by the formula of
Raiteri et al.~(1996) which is a fitting function of Padova stellar
evolutionary tracks (Bertelli et al.~1994). This formula is a function
of stellar mass $m$ and metallicity $Z$. However, the $Z$-dependence
is weak. Thus, we neglect it (we always set $Z=Z_\odot$ in the
formula).

\subsection{Stellar yields of `metal' and dust}

When stars die, they eject substantial mass of metal and dust into the
ISM. The term driving the time evolution of metal mass given by
equation (3) is the metal supplying rate, $Y_Z$, called metal
yield. Using the IMF, $\phi(m)$, and the star formation rate, $S(t)$, we
can express the metal yield as
\begin{equation}
 Y_Z(t) = \int_{m_{\rm lf}(t)}^{m_{\rm up}} 
  m_Z(m,Z[t'])\phi(m) S(t') dm\,,
\end{equation}
where $m_Z$ is the metal mass ejected from a star with mass $m$ and
metallicity $Z$, and $t'$ is given by equation (9).

The dust supplying rate, $Y_{\rm d}$, called dust yield can be
expressed likewise: 
\begin{equation}
 Y_{\rm d}(t) = \int_{m_{\rm lf}(t)}^{m_{\rm up}} 
  m_{\rm d}(m,Z[t'])\phi(m) S(t') dm\,,
\end{equation}
where $m_{\rm d}$ is the dust mass ejected from a star with mass $m$ and
metallicity $Z$, and $t'$ is given by equation (9).

\subsection{Dust destruction by supernova blast waves}

Dust grains are destroyed by SN shock waves due to shattering and
sputtering (e.g., Dwek \& Arendt 1992). This dust destruction are
observed in some SNRs as described in \S1. In this paper, we adopt the
dust destruction rate by SNe proposed by Dwek \& Scalo (1980) and 
McKee (1989): 
\begin{equation}
 D_{\rm SN}(t) = \frac{M_{\rm d}(t)}{\tau_{\rm SN}(t)}\,,
\end{equation}
where the destruction time-scale $\tau_{\rm SN}$ is defined as the
time-scale which all the ISM is swept by `dust destructive' shock
waves:  
\begin{equation}
 \tau_{\rm SN}(t) = \frac{M_{\rm ISM}(t)}{\epsilon m_{\rm SN} R_{\rm SN}(t)}\,,
\end{equation}
where $R_{\rm SN}$ is the SN occurrence rate, $m_{\rm SN}$ is the mass
swept by a single SN, and $\epsilon$ is the efficiency of the dust
destruction. The SN occurrence rate is given by 
\begin{equation}
 R_{\rm SN}(t) = \int_{8M_\odot}^{40M_\odot} \phi(m) S(t') dm\,,
\end{equation}
where we have assumed the mass range for SNe to be 8--40 $M_\odot$
(Heager et al.~2003) and $t'$ is in equation (9). If 
$t<\tau_{\rm  lf}(40\,M_\odot)$, $R_{\rm SN}=0$.
Note that we consider only Type II SNe and neglect Type Ia SNe. The
reason is discussed in \S3.

The effective mass swept by dust destructive shock wave, 
$\epsilon m_{\rm SN}$ is the important parameter. It is estimated to be 
$\sim1000$ $M_\odot$, namely $\epsilon\sim0.1$ and $m_{\rm SN}\sim10^4$ 
$M_\odot$ (McKee 1989, Nozawa et al.~2006). Recent models for
starburst galaxies in the early Universe often assume an effective mass
of $\epsilon m_{\rm SN}\sim100$ $M_\odot$ which is a factor of 10
smaller than our fiducial value (Dwek et al.~2007; Pipino et
al.~2011; Gall et al.~2011a). Their argument is that starburst
activity produces multiple SNe which make the ISM highly
inhomogeneous and the dust destruction efficiency decreases in such
medium. However, the solar neighborhood is not the case, and thus, we
keep $\epsilon m_{\rm SN}\sim1000$ $M_\odot$.

\subsection{Dust growth by `metal' accretion in the ISM}

In the ISM, atoms of some refractory elements (or refractory molecules)
may accrete onto a dust grain and may become a part of the grain. We
call this process the accretion growth of dust in the ISM (Draine
1990). Note that this process does not need nucleation, and thus, can
occur even in the ISM. A simple estimate of the growth rate is (e.g.,
Hirashita 2000) 
\begin{equation}
 G_{\rm ac}(t) = X_{\rm cold} N_{\rm d}(t) 
  \pi a^2 s_Z v_Z \rho_Z^{\rm gas}(t) \,,
\end{equation}
where $X_{\rm cold}N_{\rm d}$ is the number of dust grains in cold dense
clouds, $a$ is the grain radius, $s_Z$ is the sticking probability of
accreting metals (atoms or molecules), $v_Z$ is the thermal velocity of
the accreting metals and $\rho_Z^{\rm gas}$ is the mass density of the
accreting metals in the gas-phase. Note that all the quantities except
for $X_{\rm cold}N_{\rm d}$ in equation (15) are typical (or effective)
values averaged over various grain radii, elements, and ISM phases. The
gas-phase metal density is reduced to 
$\rho_Z^{\rm gas}=\rho_{\rm ISM}^{\rm eff}Z(1-\delta)$, where
$\rho_{\rm ISM}^{\rm eff}$ is an effective ISM mass density. We define
it as a mass-weighted average density of various ISM phases, and then,
it is determined by the density of dense molecular clouds where the
dust growth occurs. Note that $\delta=M_{\rm d}/M_Z$, the dust-to-metal
mass ratio (the dust depletion factor is $1-\delta$). For spherical
grains, $N_{\rm d}=3M_{\rm d}/(4\pi a^3 \sigma$), where $\sigma$ is the
typical material density of grains.

Equation (15) can be reduced to 
\begin{equation}
 G_{\rm ac}(t) = \frac{M_{\rm d}(t)}{\tau_{\rm ac}(t)}\,.
\end{equation}
The accretion growth time-scale $\tau_{\rm ac}$ is 
\begin{equation}
 \tau_{\rm ac}(t) = \frac{\tau_{\rm ac,0}}{Z(t) (1-\delta[t])}\,,
\end{equation}
where the normalization $\tau_{\rm ac,0}$ is the parameter determining
the process:
\begin{equation}
 \tau_{\rm ac,0}=\frac{4a \sigma}
  {3X_{\rm cold}s_Z v_Z \rho_{\rm ISM}^{\rm eff}}\,.
\end{equation}

This time-scale is very uncertain, but we will obtain 
$\tau_{\rm ac,0}=3\times10^6$ yr as the fiducial value in \S4.2 in order
to reproduce the dust-to-metal ratio at the solar neighborhood with the
SN destruction efficiency of $\epsilon m_{\rm SN}=1000$ $M_\odot$. This
value can be obtained with a set of parameters of $a=0.1$ $\mu$m
(typical size in the ISM of the Milky Way), $\sigma=3$ g cm$^{-3}$
(compact silicates), $s_Z=1$, $v_Z=0.2$ km s$^{-1}$ ($^{56}$Fe as
an accreting metal atom and thermal temperature of 100 K), 
$\rho_{\rm ISM}^{\rm eff}=1\times10^{-22}$ g cm$^{-3}$, 
and $X_{\rm cold}=0.2$. This set is just an example but ensures
that the time-scale is not outrageous.

There is a discussion about the lifetime of dense clouds (or
recycling time-scale of dense gas) should be longer than the accretion
growth time-scale for an efficient dust growth (Zhukovska et al.~2008;
Dwek \& Cherchneff 2011). According to these authors, the lifetime is
long enough to realize an efficient dust growth in the Milky Way and
even in starburst in the early Universe. Another issue is the effect of
grain size distribution which is discussed in Hirashita (2011).

\section{Stellar remnant and `metal' and dust yields}

In this section, we present new simple formulas to describe the stellar
remnant mass and yields of metal and dust which are useful to input 
into chemical evolution codes. We represent all elements heavier
than helium as just a `metal' in the formulas for simplicity, 
while yields of various elements are presented in literature. We
consider three types of stellar death: white dwarfs through the AGB 
phase, core-collapse Type II SNe, and direct collapse toward black-hole
called `collapser' (Heger et al.~2003). In this paper, we assume the
mass range for the SNe to be 8--40 $M_\odot$ (Heger et
al.~2003). The stars with mass below or above this mass range become AGB
stars or `collapsers', respectively.

We neglect Type Ia SNe for simplicity. This population of SNe is
the major source of iron element (Iwamoto et al.~1999) and may be the
source of iron dust (Calura et al.~2008). However, in respect of the
total stardust mass budget, the contribution relative to SNe II is
always less than 1--10\% (Zhukovska et al.~2008; Pipino et al.~2011). 
Since we are dealing with metal and dust as each a single component, we
safely neglect the contribution of SNe Ia.

The remnant mass, $w(m,Z)$, is taken from model calculations of AGB
stars (Karakas 2010) and SNe (Nomoto et al.~2006). Figure~1 shows the
remnant mass fraction relative to the initial stellar mass, $w/m$. This
depends on metallicity $Z$ because $Z$ in the stellar atmosphere
determines radiation pressure through opacity and the strength of
the stellar wind in the course of the stellar evolution, and affects the
remnant mass. However, as shown in Figure~1, the dependence is weak, so
we neglect it. We obtain the following fitting formula:
\begin{equation}
 \frac{w(m)}{m}=\cases{
  1 & ($m > 40\,M_\odot$) \cr
  0.13 \left(\frac{m}{8\,M_\odot}\right)^{-0.5} 
  & ($8\,M_\odot \leq m \leq 40\,M_\odot$) \cr
  0.13 \left(\frac{m}{8\,M_\odot}\right)^{-0.7}
  & ($m < 8\,M_\odot$) 
  }\,,
\end{equation}
which is shown by the solid line in Figure~1. This fitting formula
agrees with the values in Table~1 of Morgan \& Edmunds (2003) within a
$<20\%$ difference, except for $m=9$ $M_\odot$ case in which our
estimate is a factor of 2 lower than that of Morgan \& Edmunds (2003).

\begin{figure}[t]
 \centerline{\includegraphics[width=7.0cm,clip]{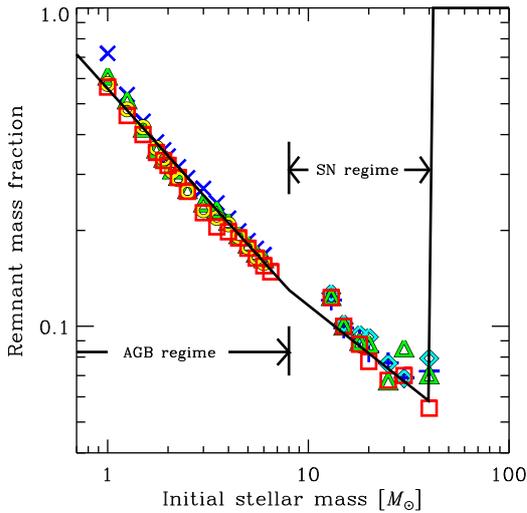}}
 \caption{Remnant mass fraction, $w/m$, as a function of the initial
 stellar mass, $m$. The data of AGB stars (1--8 $M_\odot$) are taken
 from Karakas (2010) and those of SNe (8--40 $M_\odot$) are taken from
 Nomoto et al.~(2006). The different symbols indicate different
 metallicity $Z$: $Z=0$ (plus), $Z=0.0001$ (cross), $Z=0.001$ (diamond),
 $Z=0.004$ (triangle), $Z=0.008$ (circle), and $Z=0.02$ (square). The
 solid line is a fitting function given by equation (19).}
\end{figure}

\begin{figure}[t]
 \centerline{\includegraphics[width=7.0cm,clip]{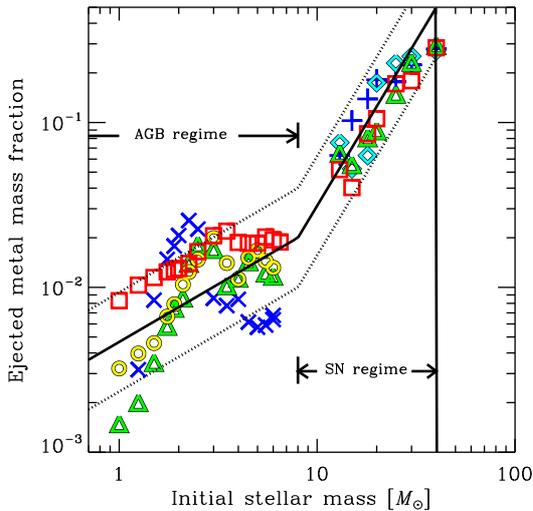}}
 \caption{Ejected metal mass fraction, $m_Z/m$, as a function of the
 initial stellar mass, $m$. The data of AGB stars (1--8 $M_\odot$) are
 taken from Karakas (2010) and those of SNe (8--40 $M_\odot$) are taken
 from Nomoto et al.~(2006). The meaning of the symbols are the same
 as in Fig.~1. The solid line is a fitting function given by equation
 (20). The dotted lines are the cases a factor of two higher or lower
 than the solid line.}
\end{figure}

For the metal yield, $m_Z$, we adopt the data taken from model
calculations of AGBs (Karakas 2010) and SNe (Nomoto et al.~2006). 
Figure~2 shows $m_Z$ relative to the initial stellar mass $m$ as a
function of $m$. While the expected $m_Z$ depends on mass $m$ and
metallicity $Z$ by a complex way, we approximate the data with a simple
power-law of only $m$ as 
\begin{equation}
 \frac{m_Z(m)}{m}=\cases{
  0 & ($m > 40\,M_\odot$) \cr
  f_Z \left(\frac{m}{8\,M_\odot}\right)^{2} 
  & ($8\,M_\odot \leq m \leq 40\,M_\odot$) \cr
  f_Z \left(\frac{m}{8\,M_\odot}\right)^{0.7}
  & ($m < 8\,M_\odot$) 
  }\,.
\end{equation}
When the normalization $f_Z=0.02$, equation (20) is the solid line in
Figure~2. As shown in the figure, the uncertainty of equation (20) is a
factor of $\sim2$. This fitting agrees with the values in Table~1
of Morgan \& Edmunds (2003) within a factor of 2 difference in the SN
regime. However, in the AGB regime, the difference is as large as the
model results by Karakas (2010). The effect of this large uncertainty of
the yield is discussed in \S6.1.

\begin{figure}[t]
 \centerline{\includegraphics[width=7.0cm,clip]{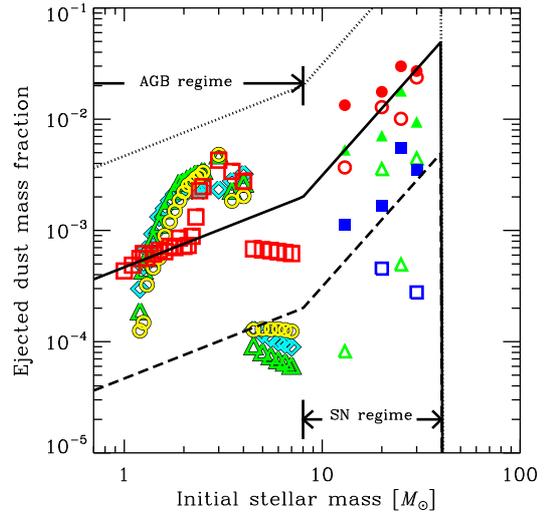}}
 \caption{Ejected dust mass fraction, $m_{\rm d}/m$, as a function of
 the initial stellar mass, $m$. The data of AGB stars (1--8 $M_\odot$)
 are taken from Zhukovska et al.~(2008) and those of SNe (8--40
 $M_\odot$) are taken from Nozawa et al.~(2007). For the AGB data, the
 different symbols indicate different  metallicity $Z$: $Z=0.001$
 (diamond), $Z=0.004$ (triangle), $Z=0.008$ (circle), and $Z=0.02$
 (square). For the SNe data, the different symbols indicate different
 ambient hydrogen density $n_{\rm H}=0.1$ cm$^{-3}$ (circle), 
 $n_{\rm H}=1$ cm$^{-3}$ (triangle), and $n_{\rm H}=10$ cm$^{-3}$
 (square). The open and filled symbols correspond to `mixed' or
 `unmixed' cases of Nozawa et al.~(2007), respectively. The dotted,
 solid, and dashed lines are the cases of $\xi=1$, 0.1, and 0.001,
 respectively, in equation (21).}
\end{figure}

The dust yield, $m_{\rm d}$, calculated by Zhukovska et al.~(2008) and
Ferrarotti \& Gail (2006) for AGBs and Nozawa et al.~(2007) for SNe are
shown in Figure~3. These yields are theoretical ones and do not
seem to be compared with observations very much yet. 
As found in Figure~3, $m_{\rm d}$ depends on mass $m$ and metallicity
$Z$ by a complex way as the metal yield $m_Z$ does. 
Moreover, the dust production by SNe is further complex because the
reverse shock moving in the ejecta of a SN may destroy the dust
produced in the ejecta (Bianchi \& Schneider 2007, Nozawa et al.~2007,
Nath et al.~2008, Silvia et al.~2010). 
This self-destruction depends on the material strength against the 
destruction\footnote{$^3$The micro-process of the destruction considered
in Nozawa et al.~(2007) is sputtering by hot gas.} and the ambient gas
density which determines the strength of the reverse shock. According to
Nozawa et al.~(2007), we plot three cases of the ambient density and
`mixed' and `unmixed' dust productions\footnote{$^4$The `mixed' and
`unmixed' mean the elemental mixing in the SN ejecta (Nozawa et
al.~2003). In the `mixed' case, there is no layer where C is more
abundant than O, then, only silicate, troilite, and corundum grains can
be formed. On the other hand, the `unmixed' case has a C-rich layer and
Fe layer and can form carbon and iron grains as well as silicate.} in
Figure~3. We adopt a simple formula for $m_{\rm d}$ as 
\begin{equation}
 m_{\rm d}(m) = \xi m_Z(m)\,,
\end{equation}
where $\xi$ is a scaling factor and means an efficiency of condensation
of metal elements. In Figure~3, we show three cases of $\xi=1$ (all
metal condenses into dust: an extreme but unrealistic case), 0.1
(fiducial case), and 0.01 (a lower efficiency case). The reader may be
anxious about a large uncertainty of this approximation. However, the
dust mass in galaxies does not depend on $m_{\rm d}$ after the 
accretion growth becomes active. This is because the growth of dust is
the dominant process of dust production after the activation as shown
later in \S6.1.

\section{Milky Way analog}

Let us calibrate parameters in the chemical and dust evolution model of
galaxies so as to reproduce the properties at the solar neighborhood in
the Milky Way. There are two parameters in the chemical evolution part:
the time-scales of star formation, $\tau_{\rm SF}$, and infall, 
$\tau_{\rm in}$. There are additional two parameters in the dust content
evolution: the time-scale of the ISM accretion growth,  $\tau_{\rm ac,0}$,
and the efficiency of the dust destruction, $\epsilon m_{\rm SN}$. In
addition, there are two parameters as uncertainties of metal and dust
yields, $f_Z$ and $\xi$. Table~1 is a summary of these parameters and
values. 

Note that we do not apply any statistical method to justify the goodness
of the reproduction of the observational constraints throughout this
paper because our aim is not to find the best fit solution for the
constraints but to demonstrate the dust content evolution in galaxies
qualitatively. This is partly due to the weakness of the observational
constraints and due to large uncertainties of the dust physics itself.

\begin{table*}[t]
\renewcommand{\arraystretch}{1.2}
\vspace{-.3cm}
\caption{Parameters and values for the solar neighborhood.}
\vspace{-.1cm}
\begin{center}
\begin{tabular}{lll}
 \hline
 Parameter & Fiducial value & Considered values \\
 \hline
 ($\tau_{\rm SF}$/Gyr, $\tau_{\rm in}$/Gyr) & (3, 15) 
     & (1, 50), (2, 20), (3, 15), and (5, 10) \\
 \hline
 ($\tau_{\rm ac,0}$/Myr, $\epsilon m_{\rm SN}$/$10^3M_\odot$) & 
     (3, 1) & (1.5, 1), (1.5, 2), (3, 0.5), (3, 1), (3, 2), (6, 0.5),
	 and (6, 1) \\
 \hline
 $f_Z$ & 0.02 & 0.01, 0.02, and 0.04 \\
 $\xi$ & 0.1 & 0.01, 0.1, and 1 \\
 \hline
\end{tabular}
\end{center}
\end{table*}%

\subsection{Chemical evolution at the solar neighborhood}

\begin{figure}[t]
 \centerline{\includegraphics[width=7.0cm,clip]{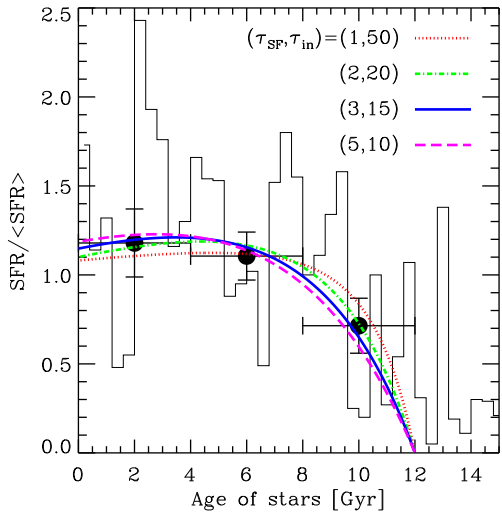}}
 \caption{Star formation history (time evolution of star formation rate)
 normalized by the average rate. The histogram is the observed history
 at the solar neighborhood reported by Rocha-Pinto et al.~(2000a). The
 filled circles with error-bars are the average of the histogram over 4
 Gyr interval and corrected by the average rate only for the age less
 than 12 Gyr which is the present age of the Milky Way assumed in
 this paper. The vertical error-bars are the standard error of the
 mean. The four lines correspond to the time evolutions with four
 different sets of time-scales of star formation and infall 
 ($\tau_{\rm SF}$/Gyr, $\tau_{\rm in}$/Gyr) as indicated in the
 panel.} 
\end{figure}

We here determine the time-scales of star formation and infall in
the chemical evolution part. First, we constrain these time-scales by
using the star formation history at the solar neighborhood reported by
Rocha-Pinto et al.~(2000a). Such a method was adopted by Takeuchi \&
Hirashita (2000). Rocha-Pinto et al.~(2000a) derived the star formation
history from the age distribution of 552 late-type dwarf stars at the
solar neighborhood. The histogram in Figure~4 is their result and shows
very stochastic nature of the history. However, our model can treat only
a smooth history. Thus, we smoothed the stochastic history by averaging
with 4 Gyr interval. The filled circles are the result. The vertical
error-bars indicate the standard error of the mean. The average history
is re-normalized by the average star formation rate for the stellar age
less than 12 Gyr which is the assumed age of the Milky Way in this
paper, although this choice of the age is arbitrary. We have tried four
cases of $\tau_{\rm SF}$ in this paper: 1, 2, 3, and 5 Gyr which
are the observed range of the time-scale (or gas consumption time-scale)
for disk galaxies like the Milky Way (e.g., Larson, Tinsley, \& Caldwell
1980). For each $\tau_{\rm SF}$, we have found $\tau_{\rm in}$ with
which we can reproduce the smoothed history as shown in Figure~4. 

\begin{figure}[t]
 \centerline{\includegraphics[width=7.0cm,clip]{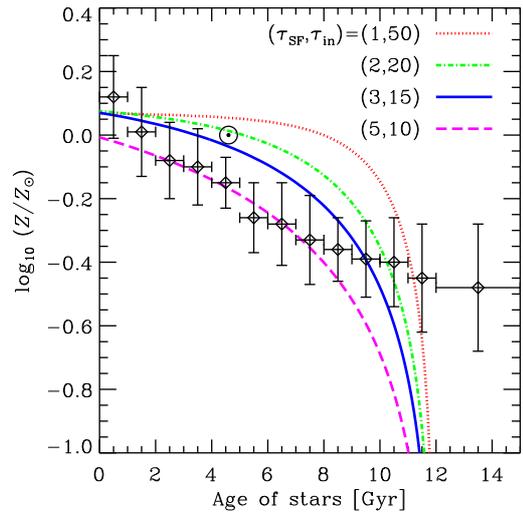}}
 \caption{Age-metallicity relation of stars. The diamonds with
 error-bars are the data of stars at the solar neighborhood reported by
 Rocha-Pinto et al.~(2000b). The solar mark ($\odot$) indicates the 
 position of the Sun on this plot. The four lines correspond to the
 model relations same as Fig.~4. The present age of the Milky Way is
 assumed to be 12 Gyr.}
\end{figure}

Next, we adopt the observed relation between the stellar age and
metallicity, so-called the age-metallicity relation, reported by
Rocha-Pinto et al.~(2000b) to further constrain 
$(\tau_{\rm SF},\tau_{\rm in})$. Rocha-Pinto et al.~(2000b) derived the
relation from the same 552 stars as Rocha-Pinto et al.~(2000a).
Their result is shown in Figure~5 by diamonds with error-bars. After
comparing with our four model lines, we have found that 
$(\tau_{\rm SF}/{\rm Gyr},\tau_{\rm in}/{\rm Gyr})=(5,10)$ case seems
the best match with the observed relation but 
$(\tau_{\rm SF}/{\rm Gyr},\tau_{\rm in}/{\rm Gyr})=(3,15)$ case is
also acceptable.

\begin{figure}[t]
 \centerline{\includegraphics[width=7.0cm,clip]{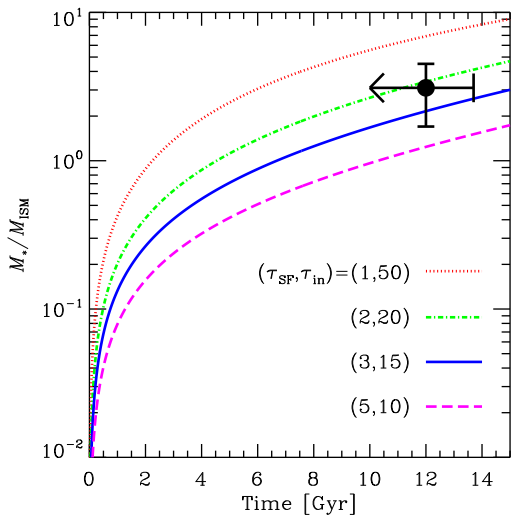}}
 \caption{Time evolution of the stellar mass relative to the ISM
 mass. The filled circle with error-bars is an estimate at the solar
 neighborhood taken from Naab \& Ostriker (2006) and references
 therein. Note that the stellar mass includes mass of remnants.
 The present age of the Milky Way should be less than the age of
 the Universe (13.7 Gyr). The four lines correspond to the model
 evolutions same as Fig.~4.}
\end{figure}

Finally, we adopt another constraint: the current stellar mass relative
to the ISM mass. Naab \& Ostriker (2006) compiled observational
constraints for the solar neighborhood. From the compilation, we adopt
the ratio of the stellar mass to the ISM mass at the present epoch of
$3.1\pm1.4$. Note that the stellar mass includes the remnant mass
(i.e. white dwarf, neutron stars, and black-holes). Figure~6 shows the
comparison of the ratio with our four star formation histories. We have
found that the two sets of 
$(\tau_{\rm SF}/{\rm Gyr},\tau_{\rm in}/{\rm Gyr})=(2,20)$ and
(3,15) are consistent with the data.

From these three comparisons, we finally adopt the case of 
$(\tau_{\rm SF}/{\rm Gyr},\tau_{\rm in}/{\rm Gyr})=(3,15)$ as the
fiducial set for the Milky Way (or more precisely, for the solar
neighborhood) in this paper.

\subsection{Dust content evolution at the solar neighborhood}

\begin{figure}[t]
 \centerline{\includegraphics[width=7.0cm,clip]{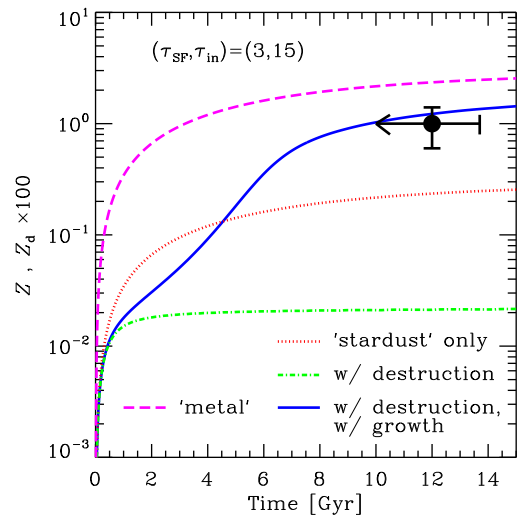}}
 \caption{Time evolution of metallicity (metal mass fraction in the ISM;
 dashed line) and dust-to-gas mass ratio (other lines) for the fiducial
 case (Table~1). The dotted line corresponds to the case only with
 `stardust' production and destruction by star formation
 (i.e. astration). The dot-dashed line correspond to the case with the
 dust destruction by SNe but without the dust growth in the ISM. The
 solid line is the case with all the processes. The filled circle
 with error-bars is an estimate of the dust-to-gas mass ratio at the
 solar neighborhood (see text). The present age of the Milky Way should
 be less than the age of the Universe (13.7 Gyr).}
\end{figure}

Here we examine the dust content evolution. First, we show the
significant effect of the dust destruction and the ISM growth. Figure~7
shows the time evolution of metallicity and dust-to-gas mass ratio for
the fiducial set of $\tau_{\rm SF}$ and $\tau_{\rm in}$ obtained in the
previous subsection. The model curves of the dust-to-gas ratio
(dotted, dot-dashed, and solid lines) are compared with the filled
circle with error-bars which is an observational estimate at the solar
neighborhood. This is obtained from metallicity $Z\approx Z_\odot$ (van
den Bergh 2000; see also Rocha-Pinto et al.~2000b) and dust-to-metal
mass ratio $\delta\approx0.5$ (Kimura et al.~2003; see below) and the
uncertainty is the quadrature of uncertainties of 30\%\footnote{ $^5$The
difference between $Z_\odot$s by Anders \& Grevesse (1989) and Asplund
et al.~(2009) accounts for the uncertainty.} in $Z$ and 20\% in
$\delta$.

If there is neither destruction nor accretion growth of dust, the
dust-to-gas ratio evolution is just the metallicity evolution multiplied
by the condensation efficiency of stardust, $\xi$, as shown by the
dotted line. We have assumed $\xi=0.1$ for the line. Once the SN
destruction of dust is turned on with a standard efficiency as 
$\epsilon m_{\rm SN}=1\times10^3\,M_\odot$ (McKee 1989;
Nozawa et al.~2006), it reduces the dust amount by a factor of ten as
shown by the dot-dashed line. This confirms that the dust destruction is
very efficient and the stardust injection is too small to compensate the
destruction (e.g., Draine 1990; Tielens 1998). Then we need the
accretion growth in the ISM to reproduce the dust-to-gas ratio
$\sim10^{-2}$ in the present Milky Way. If we assume the time-scale of
$\tau_{\rm ac,0}=3\times10^6$ yr, the dust-to-gas ratio evolution 
becomes the solid line and it reaches $\simeq10^{-2}$ which is almost
two orders of magnitude larger than the case without the growth after
several Gyr.

\begin{figure}[t]
 \centerline{\includegraphics[width=7.0cm,clip]{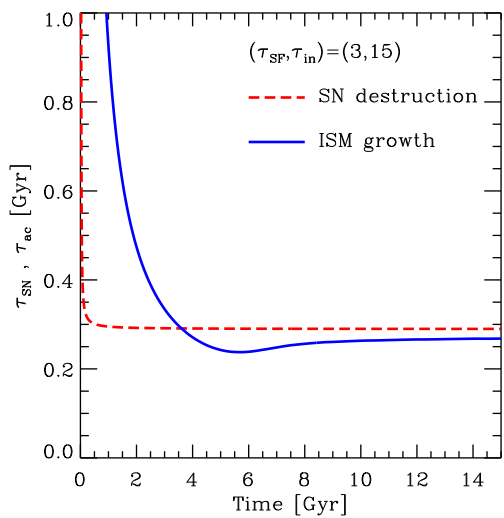}}
 \caption{Time evolution of time-scales of the dust destruction by SNe
 (dashed line) and of the dust growth in the ISM (solid line) for the
 fiducial case (Table~1).}
\end{figure}

Figure~8 shows the time evolution of $\tau_{\rm SN}$ in equation (13)
and $\tau_{\rm ac}$ in equation (17). The SN destruction time-scale
$\tau_{\rm SN}$ is almost constant promptly after the first a few
hundreds Myr. On the other hand, the accretion growth time-scale 
$\tau_{\rm ac}$ decreases gradually in the first a few Gyr. This is
because $\tau_{\rm ac}$ has a metallicity dependence as shown in
equation (17) and it decreases as the metallicity increases. At the time
around 4 Gyr, $\tau_{\rm ac}$ becomes shorter than $\tau_{\rm SN}$, and
then, the accretion growth becomes significant and the dust amount
increases rapidly. As the accretion growth proceeds, the metal abundance
in the gas phase decreases, i.e., the dust-to-metal ratio $\delta$
increases, then, $\tau_{\rm ac}$ becomes almost constant and balances
with $\tau_{\rm SN}$. We will discuss this point in \S5 more in detail.

\begin{figure}[t]
 \centerline{\includegraphics[width=7.0cm,clip]{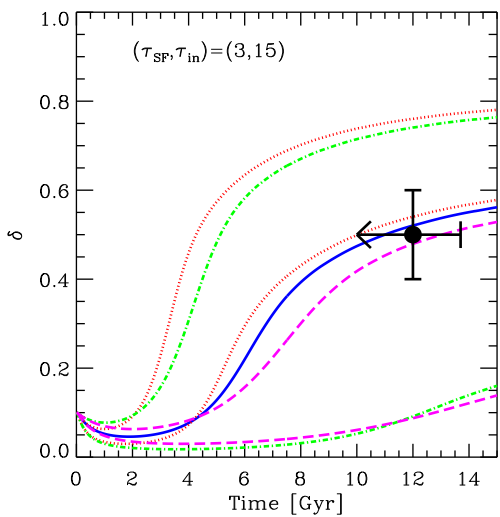}}
 \caption{Time evolution of dust-to-metal mass ratio, $\delta$, for
 various sets of the parameters of dust growth and destruction 
 $(\tau_{\rm ac,0},\epsilon m_{\rm SN})$. The solid line is the fiducial
 case with ($3\times10^6$ yr, $1\times10^3$ $M_\odot$). The two dotted
 lines correspond to two different sets of ($1.5\times10^6$ yr,
 $1\times10^3$ $M_\odot$) for the upper and ($1.5\times10^6$ yr,
 $2\times10^3$ $M_\odot$) for the lower. The two dot-dashed lines
 correspond to the sets of ($3\times10^6$ yr, $5\times10^2$ $M_\odot$)
 for the upper and ($3\times10^6$ yr, $2\times10^3$ $M_\odot$) for the
 lower. The two dashed lines correspond to the sets of ($6\times10^6$
 yr, $5\times10^2$ $M_\odot$) for the upper and ($6\times10^6$ yr,
 $1\times10^3$ $M_\odot$) for the lower. The filled circle with
 error-bars is an estimate in the Local Interstellar Cloud 
 surrounding the Sun by Kimura et al.~(2003). The present age of the
 Milky Way should be less than the age of the Universe
 (13.7 Gyr).}
\end{figure}

Figure~9 shows the time evolution of the dust-to-metal ratio, $\delta$. 
The solid line is the fiducial case which is shown in Figures~7 and 8. 
This can be compared with the observed ratio in the Local Interstellar
Cloud reported by Kimura et al.~(2003): $\delta=0.5\pm0.1$. As shown in
Figure~9, the fiducial set of $(\tau_{\rm ac,0},\epsilon m_{\rm
SN})=(3\,{\rm Myr},1\times10^3\,M_\odot)$ excellently agrees with the 
observed data. On the other hand, other sets can also reproduce the
data. For example, $(\tau_{\rm ac,0},\epsilon m_{\rm
SN})=(1.5\,{\rm Myr},2\times10^3\,M_\odot)$ or $(6\,{\rm
Myr},5\times10^2\,M_\odot)$. Interestingly, the $\delta$ evolutions
become very similar if the product of $\tau_{\rm ac,0}$ and 
$\epsilon m_{\rm SN}$ is the same. We will also discuss this point in
\S5.

\section{Determining dust-to-metal ratio}

In this section, we demonstrate the mechanism for determining the
dust-to-metal mass ratio, $\delta$, in galaxies. Starting from equations
(3) and (4), we can obtain the time evolutionary equation of
$\delta\equiv M_{\rm d}/M_Z$. Here, let us adopt the instantaneous
recycling approximation (IRA) in which we neglect the finite stellar
life-time and assume that stars with a mass larger than a certain
threshold (the turn-off mass $m_t$) die instantly when they are
formed. This approximation allows us to manage the equations
analytically and is good enough to see phenomena with a time-scale
longer than a Gyr (see Tinsley 1980 for more details). In the IRA, we
can approximate the metal and dust yields in equations (10) and (11) as 
$Y_Z\approx{\cal Y}_Z S$ and $Y_{\rm d}=\xi Y_Z\approx\xi{\cal Y}_Z S$,
where the effective metal yield 
\begin{equation}
 {\cal Y}_Z=\int_{m_t}^{m_{\rm up}} m_Z(m) \phi(m) dm 
  = 0.024 \left(\frac{f_Z}{0.02}\right)\,,
\end{equation}
where we have assumed $m_t=1$ $M\odot$. This value is not sensitive
to $m_t$. We obtain ${\cal Y}_Z=0.021(f_Z/0.02)$ if $m_t=5$ $M_\odot$.
Remembering the star formation rate $S=M_{\rm ISM}/\tau_{\rm SF}$ as in
equation (5), then, we obtain 
\begin{equation}
 \frac{1}{\delta}\frac{d\delta}{dt} \approx 
  -\frac{{\cal Y}_Z}{\tau_{\rm SF}Z}\left(1-\frac{\xi}{\delta}\right)
  -\frac{1}{\tau_{\rm SN}}-\frac{1}{\tau_{\rm ac}}\,.
\end{equation}

In the IRA, the SN destruction time-scale $\tau_{\rm SN}$ in equation
(13) can be reduced to 
\begin{equation}
 \tau_{\rm SN} \approx 
  \frac{\tau_{\rm SF}}{\epsilon m_{\rm SN} n_{\rm SN}}\,,
\end{equation}
where the effective number of SN per unit stellar mass is 
\begin{equation}
 n_{\rm SN} = \int_{8M_\odot}^{40M_\odot} \phi(m) dm = 0.010~{M_\odot}^{-1}\,.
\end{equation}
Note that $\epsilon m_{\rm SN} n_{\rm SN}$ is a non-dimensional value.
The accretion growth time-scale $\tau_{\rm ac}$ is given in equation
(17). Then, equation (23) is reduced to 
\begin{equation}
 \frac{1}{\delta}\frac{d\delta}{dt} \approx
 - \frac{\alpha+\epsilon m_{\rm SN} n_{\rm SN}}{\tau_{\rm SF}}
 + \frac{Z(1-\delta)}{\tau_{\rm ac,0}}\,,
\end{equation}
where
\begin{equation}
 \alpha = \frac{{\cal Y}_Z}{Z}\left(1-\frac{\xi}{\delta}\right)\,.
\end{equation}

In the IRA, the metallicity $Z\equiv M_Z/M_{\rm ISM}$ can be obtained
analytically (for example, see Dwek et al.~2007). Then, we have found
that $Z\to{\cal Y}_Z$ for $t\to\infty$ when $\tau_{\rm in}>\tau_{\rm SF}$.
The condensation efficiency $\xi$ is uncertain but it is of the order of
0.1 (see Figure~3). When $\delta$ is of the order of 0.1--1 as shown in
Figure~9, the ratio $\xi/\delta$ is of the order of 1 or smaller. 
Therefore, $\alpha$ is also of the order of 1 or smaller. On the other
hand, $n_{\rm SN}\sim10^{-2}$ $M_\odot^{-1}$ and 
$\epsilon m_{\rm SN}\sim10^3$ $M_\odot$, then, we obtain 
$\epsilon m_{\rm SN} n_{\rm SN} \gg \alpha$. Therefore, equation (26) is
further reduced to 
\begin{equation}
 \frac{1}{\delta}\frac{d\delta}{dt} \approx - a + b (1-\delta)\,,
\end{equation}
where $a=\epsilon m_{\rm SN} n_{\rm SN}/\tau_{\rm SF}$ and 
$b=Z/\tau_{\rm ac,0}$. If we assume $Z$ to be constant (i.e. $b$ is
constant), equation (28) can be solved analytically. The solution is 
\begin{equation}
 \delta \approx \frac{\delta_\infty\delta_0 \exp{(b-a)t}}
  {(\delta_\infty - \delta_0) + \delta_0 \exp{(b-a)t}}\,,
\end{equation}
where $\delta_0$ and $\delta_\infty$ are the values for $t=0$ and
$t\to\infty$, respectively. The asymptotic value $\delta_\infty$ for 
$t\to\infty$ is realized only when $b>a$, and is given by 
\begin{equation}
 1-\delta_\infty = \frac{a}{b} = 
  \frac{\tau_{\rm ac,0}\epsilon m_{\rm SN} n_{\rm SN}}{\tau_{\rm SF} Z}\,.
\end{equation}
This is the equilibrium value for equation (28) and we find 
\begin{eqnarray}
 1-\delta_\infty = 0.5 \left(\frac{\tau_{\rm ac,0}}{3~{\rm Myr}}\right)
  \left(\frac{\epsilon m_{\rm SN}}{10^3~M_\odot}\right)
  \left(\frac{n_{\rm SN}}{10^{-2}~M_\odot^{-1}}\right) \cr
  \times \left(\frac{3~{\rm Gyr}}{\tau_{\rm SF}}\right)
  \left(\frac{0.02}{Z}\right)\,, 
\end{eqnarray}
which excellently agrees with the results in Figure~9. 

We can fully understand the $\delta$ evolution by using equation (28).
At the beginning, the accretion term $b\sim0$ because $Z\sim0$. Then,
only the destruction term $a$ is effective. As a result, $\delta$
decreases with the time-scale of $1/a=\tau_{\rm SN}$. As $Z$ increases,
the accretion term $b$ increases and finally exceeds $a$. Then, $\delta$
increases toward $\delta_\infty$ with the evolution time-scale of
$1/(b-a)$. This decreases as $Z$ increases and $b-a$ increases. 
Therefore, the driving force of the $\delta$ evolution is $Z$. If we
call $Z$ at $b=a$ as the critical metallicity, $Z_{\rm c}$, we find 
\begin{eqnarray}
 Z_{\rm c}=\frac{\tau_{\rm ac,0}\epsilon m_{\rm SN} n_{\rm SN}}
  {\tau_{\rm SF}} =0.01~~~~~~~~~~~~~~~~~~~~~~~~~~~~~ \cr
  \times \left(\frac{\tau_{\rm ac,0}}{3~{\rm Myr}}\right)
  \left(\frac{\epsilon m_{\rm SN}}{10^3~M_\odot}\right)
  \left(\frac{n_{\rm SN}}{10^{-2}~M_\odot^{-1}}\right) 
  \left(\frac{3~{\rm Gyr}}{\tau_{\rm SF}}\right)\,.
\end{eqnarray}
When $Z>Z_{\rm c}$, the accretion growth becomes effective and $\delta$
approaches the final value $\delta_\infty$. A similar critical
metallicity has been derived by Asano et al.~(2011) with a different
way.

Equation (30) shows that the final value of $\delta$ is determined by
the equilibrium between the SN destruction and the accretion growth in
the ISM. The time-scale to reach the equilibrium is $1/(b-a)$. This is
relatively short in the fiducial case. For example, it is
0.3 Gyr when $Z=0.02$. This means that the $\delta$ evolution
proceeds with keeping the equilibrium between the SN destruction and the
accretion growth, or equivalently, $\delta=\delta_\infty$ after $Z$
exceeds $Z_{\rm c}$. This behavior is also found by the
comparison of the two time-scales, $\tau_{\rm SN}$ and $\tau_{\rm ac}$,
in Figure~8; once $\tau_{\rm ac}$ becomes shorter than 
$\tau_{\rm SN}$ at about 4 Gyr at which $Z$ exceeds $Z_{\rm c}$, 
$\tau_{\rm ac}$ turns around and approaches $\tau_{\rm SN}$ again. This
is realized by the reduction of the term $(1-\delta)$ in $\tau_{\rm ac}$
(see eq.~[17]) when $\delta$ increases from $\sim0$ to
$\delta_\infty$. Such a kind of self-regulation process determines
the dust-to-metal ratio $\delta$.

\section{Discussion}

\subsection{Effect of uncertainties of yields}

\begin{figure}[t]
 \centerline{\includegraphics[width=7.0cm,clip]{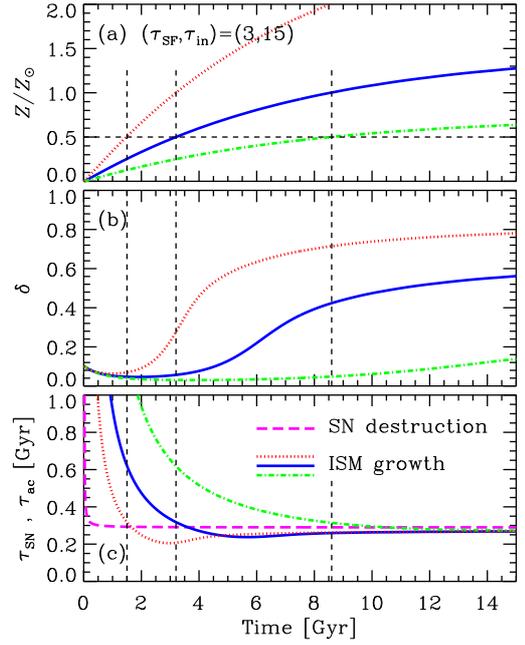}}
 \caption{Time evolution of (a) metallicity, (b) dust-to-metal mass
 ratio, and (c) time-scales of SN destruction and ISM growth for
 three different stellar metal yields: $f_Z=0.01$ (dot-dashed), 0.02
 (solid), and 0.04 (dotted). The horizontal short-dashed line in the
 panel (a) shows the critical metallicity of equation (32). The vertical
 short-dashed lines indicate the timing at which the metallicity exceeds
 the critical one for the three metal yields. The long-dashed line in
 the panel (c) is the SN destruction time-scale, but the other three
 lines are the ISM growth time-scales for the three metal yields.}
\end{figure}

Here we examine the effect of uncertainties of the normalization of
metal and dust yields. As we saw in Figure~2, our simple recipe for the
metal yield may contain a factor of 2 (or more) uncertainty. The
parameter $f_Z$ accounts for this uncertainty. In Figure~10, we show the
effect of $f_Z$. As found from the panel (a), the metallicity evolution
is scaled almost lineally by $f_Z$ as expected and the timing at
which $Z$ exceeds $Z_{\rm c}=0.5Z_\odot$ given by equation (32) for the
fiducial set of the accretion and destruction efficiencies
becomes faster as $f_Z$ is larger. From the panels (b) and (c), we find
that for each case of $f_Z$, $\delta$ increases and $\tau_{\rm ac}$
becomes shorter than $\tau_{\rm SN}$ soon after the timing for 
$Z>Z_{\rm c}$. Therefore, the timing for $\tau_{\rm ac}<\tau_{\rm SN}$,
in other words, the timing for the accretion growth activation is well
traced by $Z_{\rm c}$ in equation (32) and this is not affected by
uncertainty of $f_Z$. On the other hand, the timing for the activation
becomes faster for larger $f_Z$. The metallicity dependence on the
final value of $\delta$ is explicit as found in equation (31).

\begin{figure}[t]
 \centerline{\includegraphics[width=7.0cm,clip]{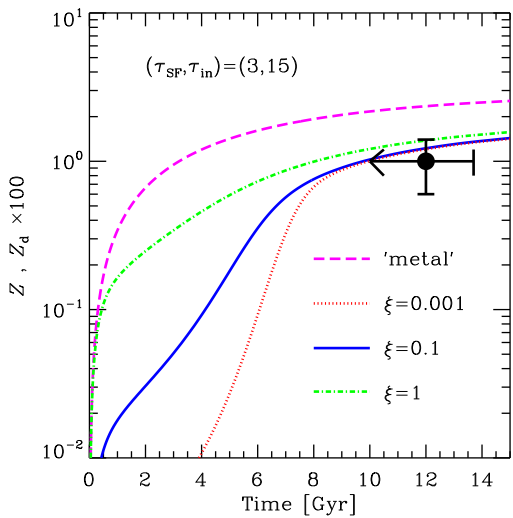}}
 \caption{Same as Fig.~7 but for different condensation efficiencies in
 the stellar ejecta as indicated in the panel.}
\end{figure}

Figure~11 shows the effect of the dust yield. As seen in Figure~3, our
recipe for the stardust yield has a factor of 10 or larger uncertainty
because of a large uncertainty in the adopted model calculations. In
Figure~11, we show the cases with a factor of 10 larger or smaller
yield than the fiducial one. Other parameters are the same as the
fiducial set, so that we have the same evolutions of the metallicity and
the time-scales of the SN destruction and accretion growth as shown by
the solid lines in Figure~10. Before the growth activation at
around 4 Gyr, the dust amounts show a large difference, however, they
converge nearly the same amount after the activation. This is because
the final value of $\delta$ given in equation (31) does not depend
on the dust yield. Therefore, we conclude that the dust content in
galaxies is independent of the stardust yield after the grain
growth in the ISM becomes active, or equivalently, the
metallicity exceeds the critical one.

\subsection{What kind of dust is formed by the ISM growth?}

We have shown that the main production channel of dust is the accretion
growth in the ISM of the present-day Milky Way. This conclusion had
been obtained also in the literature. For example, Zhukovska et
al.~(2008) argued that the mass fraction of stardusts in total
dust is only 0.1--1\% based on a more sophisticated chemical
evolution model than this paper (see their Fig.~15); more than 99\% of
dust is originated from the accretion growth in the ISM. It is
also well known that some interplanetary dust particles show a highly
enhanced abundance of deuterium and $^{15}$N relative to the solar
composition, which is a signature of the molecular cloud origin because
such isotopic fractionations are expected in low temperature environment
(e.g., Messenger 2000). Therefore, dust produced by the ISM accretion
exists. Then, we have a very important question; what kinds of dust
species are formed by the accretion growth in the ISM?

In molecular clouds, many kinds of ices such as H$_2$O, CO, CO$_2$,
CH$_3$OH have been detected (e.g., Gibb et al.~2000). These ices are
condensed onto pre-existent grains. In these ices, some chemical
reactions and ultraviolet photolysis (and cosmic rays) process the
material and may make it refractory. As a result, so-called
`core-mantle grains' coated by refractory organics would be formed
(e.g., Li \& Greenberg 1997). Indeed, such a grain has been found in
cometary dust: olivine particles produced by a Type II SN coated by
organic matter which seems to be formed in a cold molecular cloud
(Messenger et al.~2005). Therefore, the ISM dust probably has 
core-mantle or layered structures. Moreover, the composition can 
be heterogeneous: for example, graphite coated by silicate,
silicate coated by graphite, silicate coated by iron, etc. The
formation of such grains does not seem to be studied well. Much more
experimental and theoretical works are highly encouraged.

If we can find signatures of the dust accretion growth in the ISM
of galaxies by astronomical observations (i.e. very distant
remote-sensing), it proves the growth ubiquitous. A possible evidence
already obtained is a huge mass of dust in galaxies which requires the
accretion growth as discussed in this paper. It is worth studying how to
distinguish stardust grains (or grain cores) and ISM dust (or mantle) by
observations, e.g., spectropolarimetry, in future.

\subsection{Dust amount in the proto-solar nebula}

We have shown that the dust amount is very small before the ISM growth
becomes active. For example, the dust-to-gas mass ratio is of the order
of $10^{-4}$ at the first a few Gyr from the formation of the Milky Way
(or the onset of the major star formation at the solar neighborhood).
If the dust-to-gas ratio in the proto-solar nebula was $10^{-4}$, the
planet formation might be difficult. Fortunately, the activation of the
ISM growth is expected to be about 8 Gyr ago at the solar
neighborhood. Thus, it is well before the solar system formation.
Indeed, we expect the dust-to-gas ratio of several times $10^{-3}$ at
4--5 Gyr ago (see Figure~7). Moreover, the dust-to-gas ratio may be
much enhanced in the proto-solar nebula relative to the average ISM.
This is because the accretion growth is more efficient for higher
density and the density in the proto-solar nebula is several orders of
magnitude higher than that in molecular clouds. Therefore, even if the
solar system formation is before the activation of the ISM growth
globally, the dust growth may be active locally in the proto-solar
nebula. In this case, the planet formation is always possible if there
is enough metal to accrete onto the pre-existent seed grains, even
before the global growth activation. This is an interesting issue to
relate to the Galactic Habitable Zone where complex life can be formed
(Lineweaver et al.~2004). We will investigate it more in future.

\acknowledgments{
The author thanks to anonymous referees for their many suggestions which
were useful to improve the presentation and the quality of this paper.
The author is grateful to T.~Kozasa and A.~Habe for interesting
discussions and for their hospitality during my stay in Hokkaido
University, Sapporo where this work was initiated, to R.~Asano,
H.~Hirashita, and T.~T.~Takeuchi for many discussions, to T.~Nozawa for
providing his dust yields in SNe, and to H.~Kimura, the chair of
the convener of the `Cosmic Dust' session in the AOGS 2010 meeting, for
inviting me to the interesting meeting in Hyderabad, India.
This work is supported by KAKENHI (the Grant-in-Aid for Young
Scientists B: 19740108) by The Ministry of Education, Culture, Sports,
Science and Technology (MEXT) of Japan.
}


\email{A.K.Inoue (e-mail: akinoue@las.osaka-sandai.ac.jp)}
\label{finalpage}
\lastpagesettings
\end{document}